\documentclass[manuscript]{aastex}

\shorttitle{Discovery of an extended X-ray jet in AP Librae}
\shortauthors{S. Kaufmann et al.}

\begin{document}


\title{Discovery of an extended X-ray jet in AP Librae}

   \author{S.~Kaufmann}
   \affil{Landessternwarte, Universit\"at Heidelberg, K\"onigstuhl, D-69117 Heidelberg, Germany\\
       \email{S.Kaufmann@lsw.uni-heidelberg.de} }

   \author{S.J.~Wagner}
   \affil{Landessternwarte, Universit\"at Heidelberg, K\"onigstuhl, D-69117 Heidelberg, Germany\\
	}
       
   \author{O. Tibolla}
   \affil{Institut f\"ur Theoretische Physik und Astrophysik, Universit\"at W\"urzburg, Campus Hubland Nord, Emil-Fischer-Str. 31, D-97074 W\"urzburg, Germany \\
	}

\begin{abstract}
{\it Chandra} observations of the low-energy peaked BL Lac object AP Librae revealed the clear discovery of a non-thermal X-ray jet. 
AP Lib is the first low energy peaked BL Lac object with an extended non-thermal X-ray jet that shows emission into the VHE range.
The X-ray jet has an extension of $\sim 15'' (\approx 14 \; \rm{kpc})$. The X-ray jet morphology is similar to the radio jet observed with VLA at 1.36 GHz emerging in south-east direction and bends by $50^\circ$ at a distance of $12''$ towards north-east. The intensity profiles of the X-ray emission are studied consistent with those found in the radio range.
The spectral analysis reveals that the X-ray spectra of the core and jet region are both inverse Compton dominated. 
This adds to a still small sample of BL Lac objects whose X-ray jets are IC dominated and thus more similar to the high luminosity FRII sources than to the low luminosity FRI objects, which are usually considered to be the parent population of the BL Lac objects.
\end{abstract}


\keywords{BL Lac objects: individual (AP Lib), galaxies: active, galaxies: jets, X-rays: galaxies}

\section{Introduction}

The low-energy peaked BL Lac object AP Librae (AP Lib, PKS 1514-241) has a redshift of 
z=0.0486 \citep{Disney1974}
and is located at $\alpha_{\rm{J2000}} = 15^{\rm{h}} 17^{\rm{m}} 41.81313^{\rm{s}} \pm 0.00002^{\rm{s}}$, $\delta_{\rm{J2000}} = -24^\circ 22' 19.4759'' \pm 0.0003''$ as determined from VLBI observations by \cite{Lambert2009}.
It has been classified as a BL Lac object by \cite{Strittmatter1972} and \cite{Bond1973}.

AP Lib is well known as one of the most active blazars in the optical band. 
In data from 1989, intra-day variability was detected with a very high rate of change of $0.06\pm 0.01$ mag/hr \citep{Carini1991}. Even on shorter time scales of 20min, variation of up to $0.5 \; \rm{mag}$ have been detected in 1973 \citep{Miller1974}.

AP Lib was historically classified as a so-called radio selected BL Lac objects (RBL). In the 1990s, BL Lac objects were found mainly in radio or X-ray surveys and therefore classified as radio or X-ray selected BL Lac objects.
The spectral energy distributions (SED) of BL Lac objects show two prominent peaks which are commonly described in a leptonic model as synchrotron and inverse Compton (IC) emission, respectively. 
The peak energy of the low-energy (synchrotron) component is used to classify BL Lac objects
as low-energy peaked BL Lac object (LBL) and high-energy peaked (HBL), with occasional references to a group of intermediate-energy peaked BL Lac objects (IBL). This classification is frequently based on the slope of the X-ray spectrum.
\cite{Ciliegi1995} found that the X-ray spectrum of AP Lib can be described by a power law with photon index of $1.5-1.7$. 
AP Lib is thus classified as low-energy peaked BL Lac object (LBL) and it is assumed that the X-ray emission of the core is due to IC scattering of the synchrotron emission, and, possibly, external radiation.
The peak energy of the gamma-ray component of LBL objects is expected to arise in the keV-GeV range and therefore a rather low flux in the TeV $\gamma$-ray range is expected (close to or below the detection limit of current Cherenkov telescopes). Hence, it is rather unexpected to detect very high energy (VHE,E$>$ 100 GeV) $\gamma$-ray emission from an LBL. 
Therefore it was remarkable, that in June/July 2010, VHE $\gamma$-ray emission was detected from the position of AP Lib \citep{Hofmann2010}.


The standard picture of AGN, e.g. \cite{Blandford1974}, 
explains the different observational characteristics of AGN with the orientation of the jet axis to the line of sight to the observer. 
BL Lac objects are interpreted as aligned (beamed) versions of Fanaroff-Riley I (FRI, \cite{FR1974}) radio galaxies while steep and flat spectrum radio quasars (SSRQ and FSRQ) are aligned versions of FRII galaxies. 
BL Lac objects are known to have very energetic jets pointing under small viewing angle towards the observer.

Radio observations of AP Lib at 1.4 GHz 
reveal the detection of one-sided, diffuse radio emission at the arcmin scale \citep{Condon1998}. 
Observations with the Very Large Array (VLA) at 1.36 GHz and 4.6 GHz result in the detection of a one-sided radio jet emerging to the south-east direction and bending towards north-east at $\sim 12$ arcsec distance from the core \citep{Cassaro1999}. Very Long Baseling Array (VLBA) observations detected the radio jet at milli-arcsec scale which emerges to the south (MOJAVE\footnote{http://www.physics.purdue.edu/MOJAVE/}, \cite{Lister2009}).

The inhomogeneous collection of published X-ray jets (status March 2013: 113 X-ray jets), the XJET database\footnote{http://hea-www.harvard.edu/XJET/}, contains $77\%$ high luminosity sources (FRII, quasars) and $\sim 23\%$ low luminosity sources (20 FRI, 5 BL Lac objects and one Seyfert 1 galaxy). 
The X-ray emission detected from the jets of FRI sources are dominated by synchrotron emission \citep{Harris2006}. Most high luminosity sources (quasars, FRII) have X-ray jets with flat ($\alpha < 1$) spectra, suggesting this part of the spectrum to be dominated by inverse Compton emission (e.g. \cite{Harris2006}, \cite{Worrall2009}). 
While inverse Compton scattering of CMB photons is an explanation for flat X-ray spectra preferred by many, problems with this interpretation have been discussed by \cite{Harris2006} and \cite{Worrall2009}.





Among the sources in this database, only discovered in the radio galaxies Cen A and M87 and the FSRQ PKS 1222+216 and PKS 1510+089 have been traced up to TeV energies.
AP Lib is the first BL Lac object with an extended X-ray jet that has been detected in the TeV energy range.

The luminosity distance of AP Lib, 
using $H_0 = 70 \; \rm{km \; s^{-1} \; Mpc^{-1}}$, is  $d_L = 210 \; \rm{Mpc}$. 
The scale is $1'' = 0.95 \;\rm{kpc}$. 






 \begin{figure}[h]
	\includegraphics[width=\columnwidth]{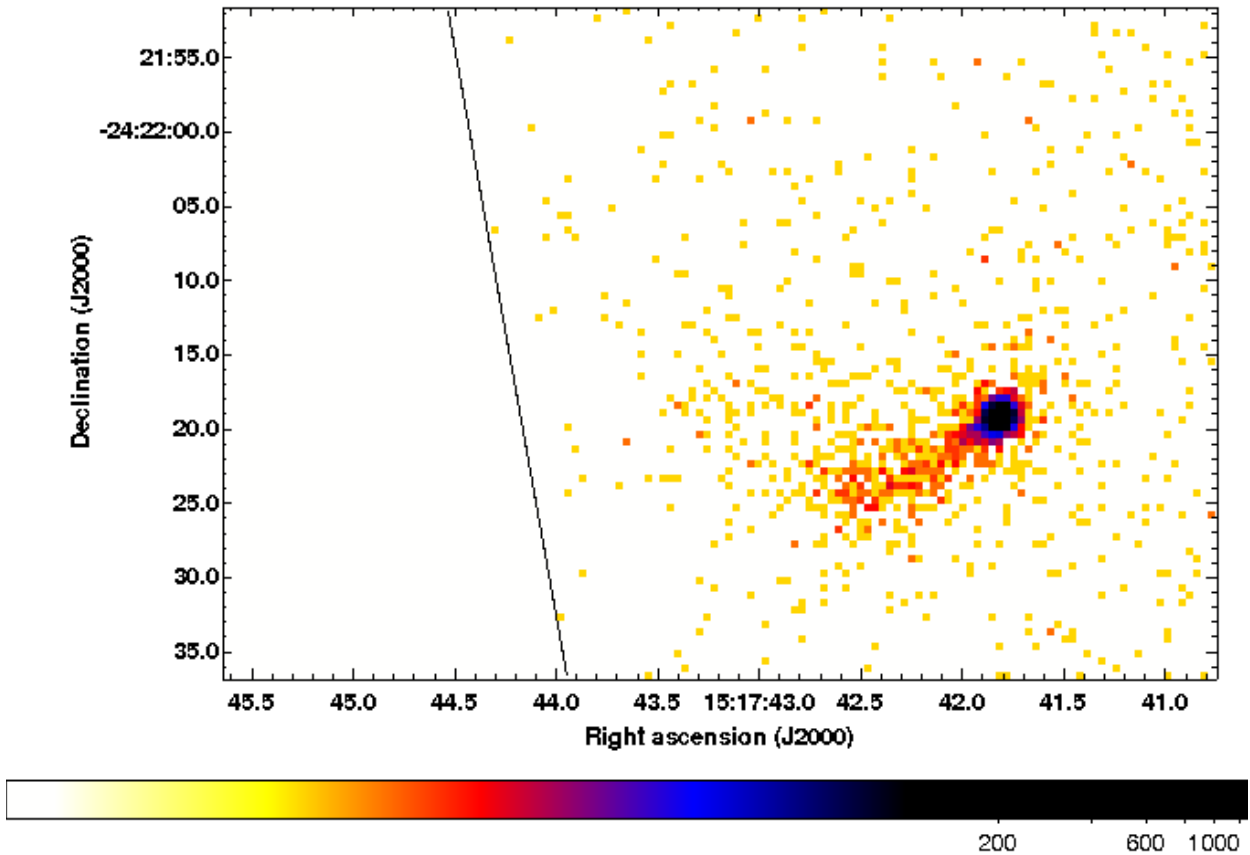}
   \caption{
X-ray count map of AP Lib from 0.2 to 8 keV extracted from the 12.8 ks observation by {\it Chandra}. The 
jet is clearly visible. Due to the used subarray, the observed frame is cut at the left part of the image, indicated by the line. 
}
   \label{1514_Xjet}
 \end{figure}

\section{Extended X-ray jet}

X-ray observations of AP Lib were analysed to search for high-energy emission of the one-sided radio jet. The source 
was observed with {\it Chandra} on July 4, 2003 (ObsID: 3971) 
with an exposure of 12.8 ks. The data were taken in timed exposure mode using a subarray of 1/8 of the chip (128 rows). This mode decreases the frame time to 0.4 s and reduces pileup.
The data were re-calibrated using the calibration database CALDB. 
The {\it Chandra} data have bee analysed with the software {\tt CIAO v4.1}.

A clear X-ray jet was detected \citep{Kaufmann2011} in this {\it Chandra} observation (see Fig. \ref{1514_Xjet}). The jet extends toward the south-east direction of AP Lib. 
Unfortunately the exposure is rather low, so that the real extension of the jet cannot be measured.

Despite the timed exposure mode, the brightness of the core causes a faint "readout artefact" visible in the column in which the bright core region is located. 
For the spectral analysis, the signal of the influenced columns is replaced by a typical background level to correct for this artefact. 
The tool {\tt acisreadcorr} has been used to correct the data. For the used subarray, the {\tt BACKSCAL} header keyword had to be modified\footnote{http://cxc.harvard.edu/ciao/threads/acisreadcorr/index.html\#subarray}.

As can be seen in Fig. \ref{1514_Xjet}, the X-ray jet emerges in the South-East direction up to an extension of $\sim 12''$ and bends with an angle of $\approx 50^\circ$ to the North-East. The jet broadens towards the outer regions and becomes fainter. 

\begin{figure}[h]
              \includegraphics[width=\columnwidth]{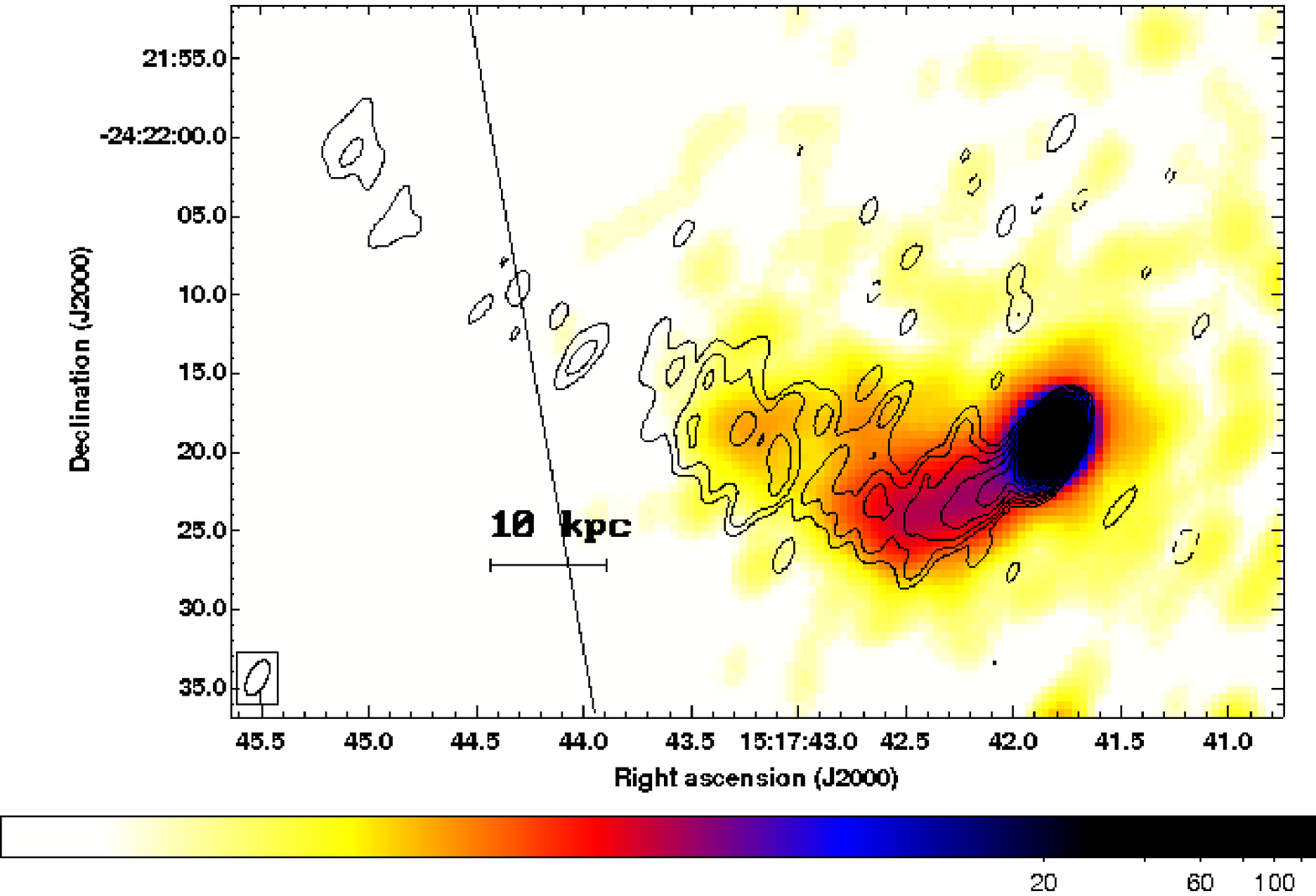}
   \caption{
X-ray count map of the energy 0.2 to 8 keV with adaptive smoothing using an elliptical Gaussian.
The X-ray jet is clearly visible. 
Contours of the VLA observation of AP Lib in A+B configuration at 1.36 GHz and the restoring beam 
in the position angle (PA) $28^\circ$ are taken from \cite{Cassaro1999}. 
}
   \label{1514_Xjet_VLA}
 \end{figure}

In order to compare the morphology of the X-ray jet to the radio jet studied with VLA by \cite{Cassaro1999}, the X-ray count map has been smoothed (see Fig. \ref{1514_Xjet_VLA}) with an elliptical Gaussian with major axis $3''$ and minor axis $2''$ using the ftool {\tt fgauss}. The shape for the elliptical Gaussian matches the beam profile
of the VLA data.
The VLA radio contours (described in section \ref{section_radio}) are overlayed onto the smoothed X-ray map (see Fig. \ref{1514_Xjet_VLA}).
The jet in the radio and X-ray bands have very similar morphologies. They extend in the same direction, bend at the same distance by the same amount, and have comparable spatial profiles.  Neither the X-ray nor the radio jet displays knots, hotspots or other features of high contrast. No feature on the counter-jet side could be detected in either band. A quantitative comparison is presented in section \ref{profiles}.

\subsection{Intensity profiles}
\label{profiles}

Radial profiles have been extracted with the tool {\tt dmextract} from the whole source (core and jet) using 20 equidistant circular annuli of width $\sim 1''$. In addition, a fraction of the annuli (wedge) with opening angle $100^\circ$ in the direction of the jet has been used. 
A circular region close to the source in North-West direction with radius $\sim 9''$ was used to determine the background for the radial profile.
The two methods (complete annuli and a fraction of the annuli) used to obtain the radial profile do not show significant differences. 
This radial profile and its comparison to the PSF are used to identify the extension of the jet and to find the best regions to obtain the core and jet spectra.



The radial profiles have been extracted for two different energy sub-bands (0.2 - 1.5 keV (S) and 1.5 - 8 keV (H)) and the hardness ratio profile has been calculated to study the spectral trends in
 the jet region. 
The hardness ratio profile has 
no significant trend 
within $15''$.

The PSF for the specific on-axis angle of the source was derived using the {\it Chandra} Ray Tracer (ChaRT \footnote{http://asc.harvard.edu/chart/}) which simulates the High Resolution Mirror Assembly (HRMA) based on an input energy spectrum of the core and the exposure of the observation. The output from ChaRT can be modeled, taking into account instrument effects of the various detectors, using the software MARX \footnote{http://space.mit.edu/CXC/MARX/} to obtain the image of the simulated PSF on the detector.  
The radial profile for this PSF has been created with wedge annuli with an opening angle of $100^\circ$.

\begin{figure}[h]
\includegraphics[width=\columnwidth]{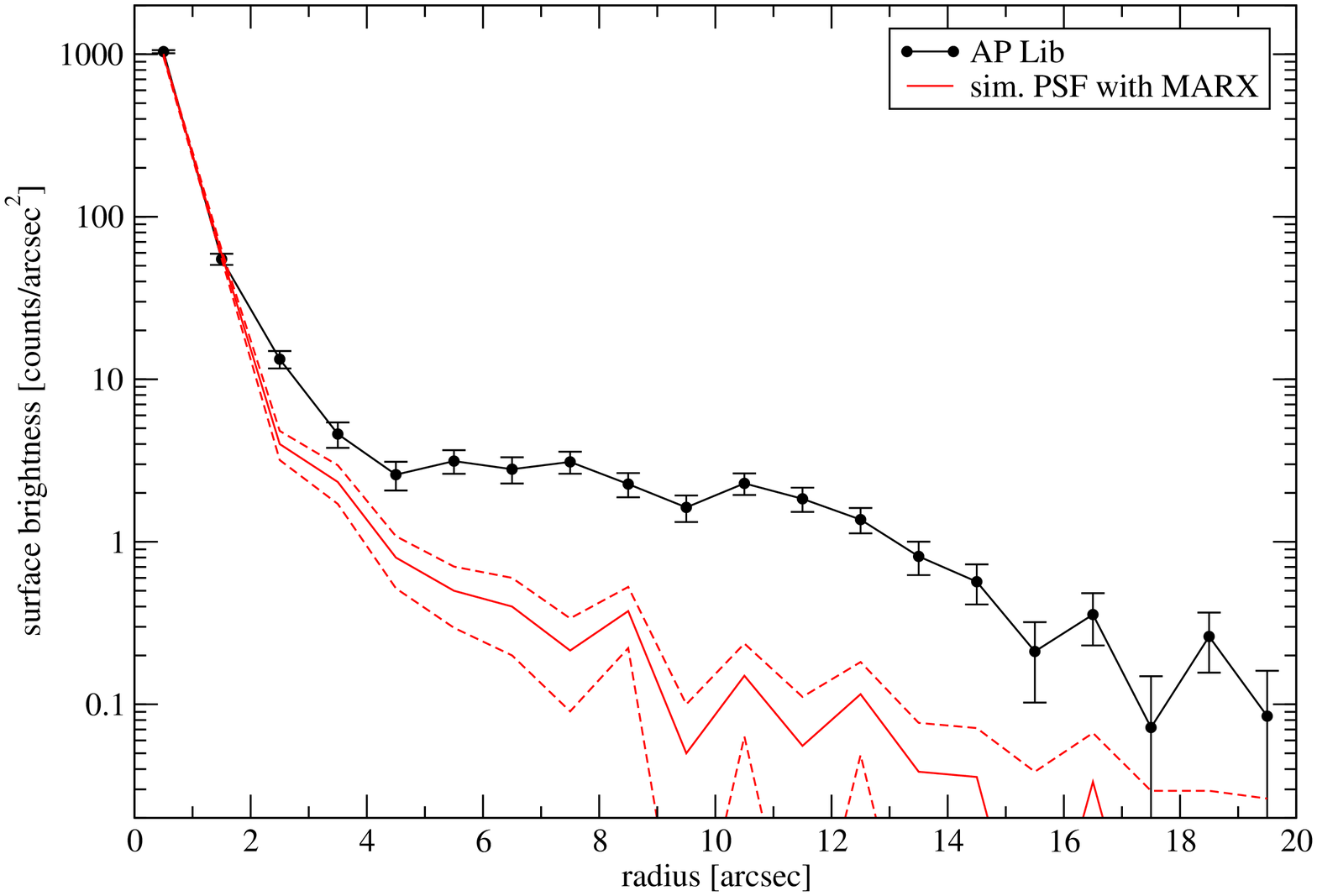}
   \caption{
Background subtracted radial profile of AP Lib, extracted from wedge annuli regions for the energy range 0.2 - 8 keV in the direction of the jet with an opening angle of $100^\circ$. The line represents the PSF simulated with MARX based on the spectrum of the core region.
The dashed lines represent the uncertainty range for the PSF.
}
\label{1514_chandra_rprofilepanda_psf}
 \end{figure}

The radial profile of the X-ray source deviates significantly from the PSF at radii $>2''$ 
(see Fig. \ref{1514_chandra_rprofilepanda_psf}).
The jet at radii $>3''$ has a linear structure and therefore the intensity profile for the jet has been created using rectangular regions along the jet.

\begin{figure*}[h]
\centerline{
\includegraphics[width=0.5\columnwidth]{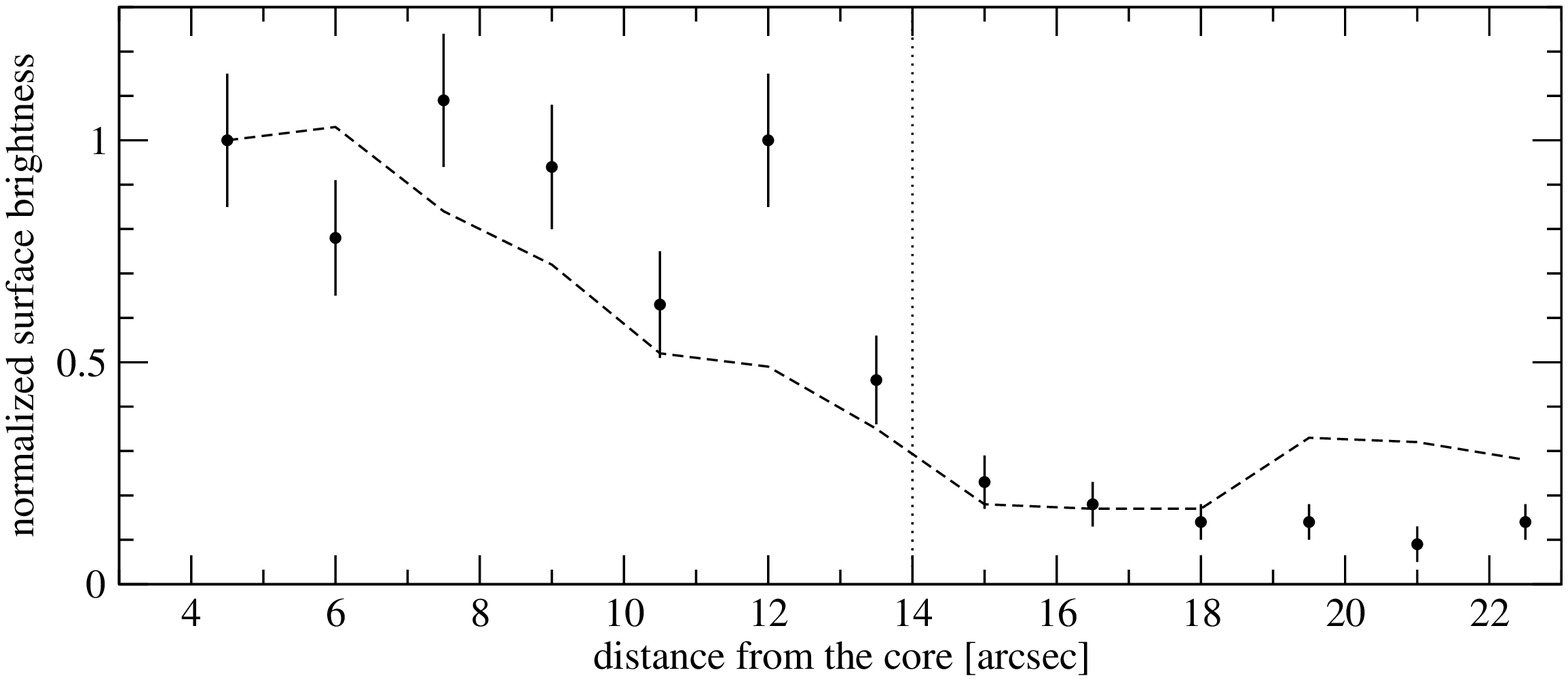}
\hfil
\includegraphics[width=0.5\columnwidth]{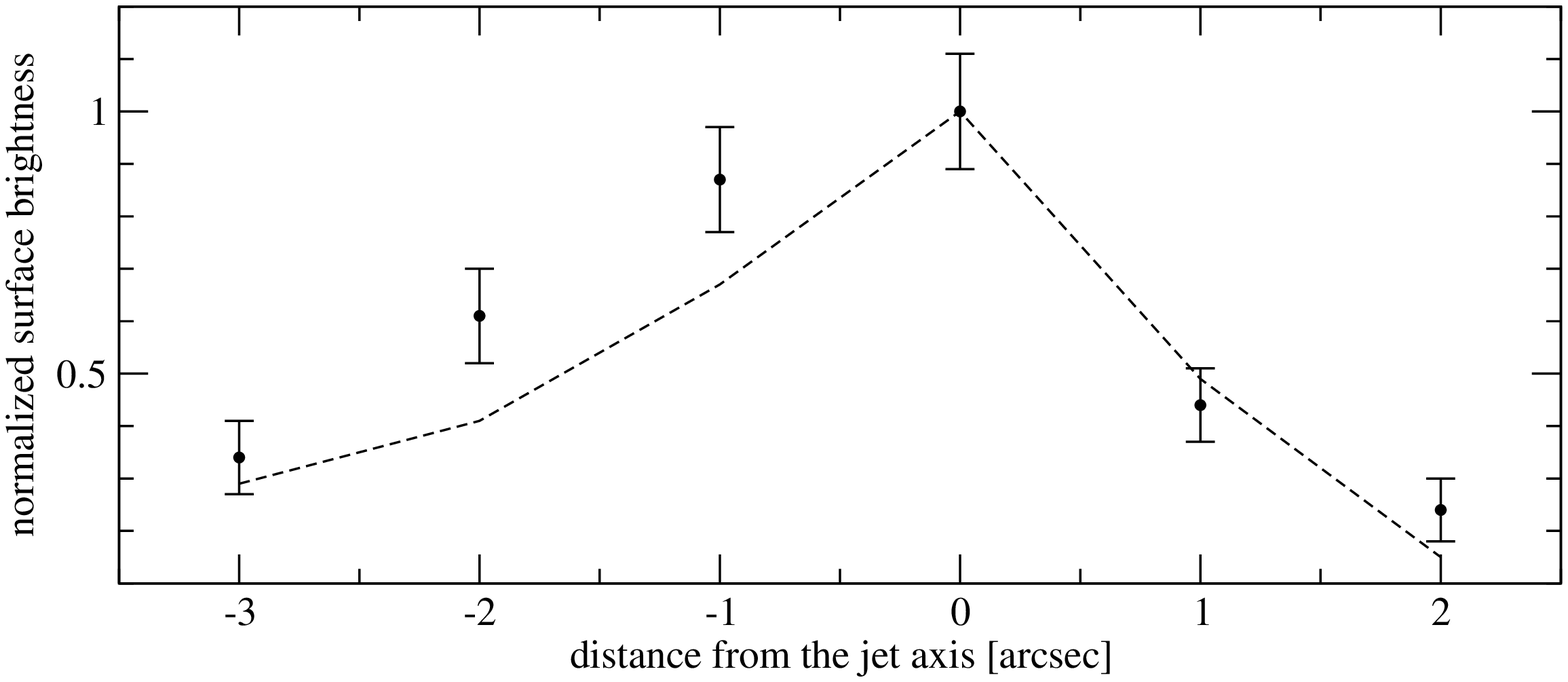}
}
   \caption{
{\it left}:
Jet intensity profile of the X-ray (circles) and radio emission (dashed line) from box regions along the jet. The brightness has been normalized to the first value with a surface brightness of $5.2 \;\rm{counts/arcsec^2}$ for the X-ray emission and $16.15 \;\rm{mJy/beam}$ for the radio emission.
The dotted vertical line indicates the region of the bending of the jet. 
{\it right}:
Transverse jet profile of the X-ray emission shown as circles and the radio emission indicated as dashed line. The brightness has been normalized to the highest value with a surface brightness of $7.6 \;\rm{counts/arcsec^2}$ for the X-ray emission and $26.1 \;\rm{mJy/beam}$ for the radio emission.
}
\label{1514_chandra_boxeslprofile}
\label{1514_chandra_boxesprofile}
 \end{figure*}

The intensity profile of the X-ray and radio jet has been extracted by integrating the counts at equidistant steps of $1.5''$ along the jet and perpendicular to the direction of the jet to compare the X-ray and radio morphology of the jet (see Fig. \ref{1514_chandra_boxesprofile}). 
In order to account for the jet bend, profiles along and across the jet are determined from $3''$ to $13.5''$ along PA = $120^\circ$ and from $13.5''$ to $22.5''$ along PA = $70^\circ$.
In order to avoid contributions from the core, a minimum distance of $3''$ was chosen. 
As can be seen in Fig. \ref{1514_chandra_boxesprofile}, the same morphology is detected of the X-ray and radio emission with a slightly shallower gradient of radio flux in the outer region of the jet.

The transversal profile (see Fig. \ref{1514_chandra_boxeslprofile}) of the jet from $3''$ up to $13.5''$ ($\approx 2.9 - 12.8 \; \rm{kpc}$) was determined
along  PA = $120^\circ$.
The transversal profile of the X-ray and radio emission is comparable and the jet width is $5'' \approx 4.8 \;\rm{kpc}$ in both energy bands.

The profiles of both energy bands are compatible. This suggests that the same particles are responsible 
for the radio and X-ray jet.

\subsection{Spectra}
\label{spectra}

The spectra of the core and the jet of AP Lib and of a background region are determined with the tool {\tt dmextract}. The response files are obtained with the tool {\tt mkrmf} and the ARF are created using {\tt asphist} and {\tt mkarf}.

\begin{figure}[h]
\includegraphics[width=\columnwidth]{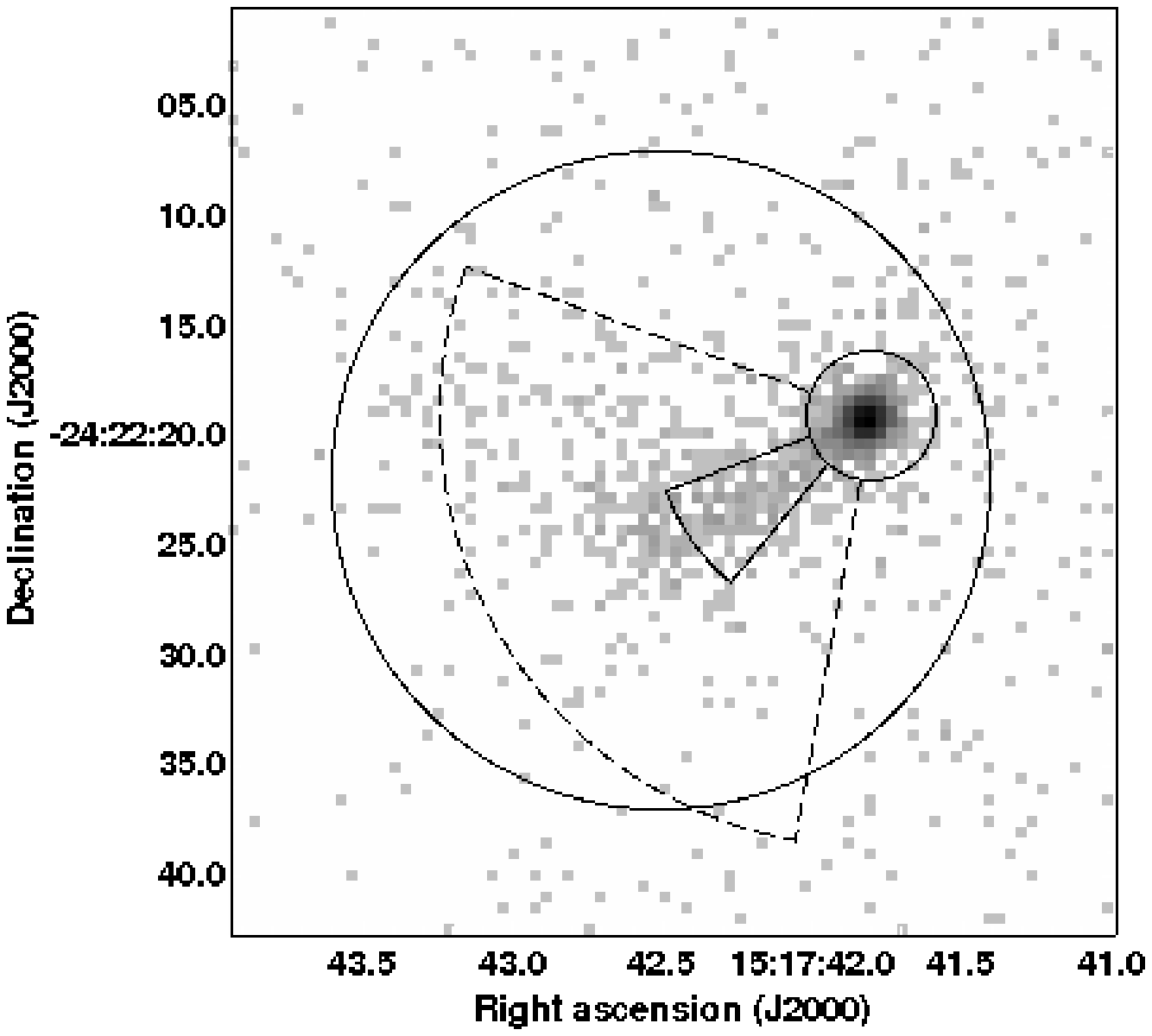}
   \caption{X-ray count map of the {\it Chandra}  observation with the different regions used for the spectral analysis of the jet.
}
   \label{1514_chandra_regions}
 \end{figure}

The X-ray spectrum of the core was extracted using a circular region with radius $3''$. This extraction radius was chosen, since the radial profile matches the simulated PSF out to this radius.
The background spectrum was determined within a region of radius $25''$ close-by to the source in the North direction.
To obtain the spectrum of the extended emission, a circular region with radius $15''$ was used in which the region around the core was excluded (see Fig. \ref{1514_chandra_regions}). 
As can be seen from the radial profile, the core exclusion region is large enough to avoid any influence of the core photons. 
Based on the radial profile of the core and jet (Fig. \ref{1514_chandra_rprofilepanda_psf}) and the shown PSF, the wing of the PSF of the core contributes $<10 \%$ to the jet spectrum obtained in the range $> 3''$.

The X-ray spectra have been binned with the tool {\tt grppha} to obtain at least 25 counts per bin to reach the necessary significance for the $\chi^2$ statistics. The program {\tt xspec v12} was used to fit the X-ray spectra in the energy range  0.2 to 8 keV. The uncertainties on the model parameters are given as confidence intervals. The fit parameter is changed 
by $\Delta \chi^2 = 2.71$. This represents the $90\%$ confidence interval.

The spectrum of the core can be well ($\chi^2/dof=165/147$) described by a power law of the form $N(E) = N_0 \times E^{-\Gamma}$ with $\Gamma = 1.58 \pm 0.04$ taking into account the Galactic absorption of $N_H = 8.36\times 10^{20} \; \rm{cm^{-2}}$ (LAB survey\footnote{http://heasarc.gsfc.nasa.gov/cgi-bin/Tools/w3nh/w3nh.pl}, \cite{Kalberla2005}). The resulting flux is $F_{\rm{core,2-10keV}} = (2.9 \pm 0.1) \times 10^{-12} \; \rm{erg \; cm^{-2} \; s^{-1}}$. 
Although no hints for pileup appear in the residuals of the power law fit, a test for pileup has been performed. Therefore, the spectrum of the core region has been extracted from an annulus region of the same size, in which the innermost pixels (inner radius of $\sim 1''$) are excluded. The spectral slope of the determined spectrum ($\Gamma = 1.6 \pm 0.1$) is comparable to the core spectrum of the circular region and therefore no pileup was detected.

To quantify and to search for spectral differences, three different regions have been used as illustrated in Fig. \ref{1514_chandra_regions}. The {\it extended emission region} is a circular region with radius $15''$ excluding the core region with radius $3''$. The {\it wide jet region} is a wedge region with radius $20''$ with opening angle  $100^\circ$ along the jet and the {\it inner jet region} is a wedge region with opening angle of $30^\circ$ and radius of $10''$. 

The spectrum of the jet obtained from the {\it extended emission region} can be described by $\Gamma = 1.8 \pm 0.1$ taking into account the Galactic absorption
($\chi^2/dof = 13/18$). The resulting flux is $F_{\rm{jet,2-10keV}} = (2.3 \pm 0.3) \times 10^{-13} \; \rm{erg \; cm^{-2} \; s^{-1}}$.

The spectrum from the {\it wide jet region} can be fit with a photon index of $\Gamma = 1.8 \pm 0.2$ ($\chi^2/dof = 6/13$) comparable to the one above. The resulting flux is $(1.6 \pm 0.3) \times 10^{-13} \; \rm{erg \; cm^{-2} \; s^{-1}}$.
An even smaller region, the  {\it inner jet region} was used to determine the spectrum of the inner parts of the jet. Since the region is very small, the spectrum consist of only a few photons and therefore a reduced flux of 
$(6.0 \pm 1.7) \times 10^{-14} \; \rm{erg \; cm^{-2} \; s^{-1}}$ 
resulted. The slope ($\Gamma = 1.9 \pm 0.5$) is comparable to the above determined spectral fits. 
For all spectral fits, the Galactic absorption was used as fixed parameter and no hint for additional absorption was found.

\begin{figure}[h]
\includegraphics[width=0.8\columnwidth]{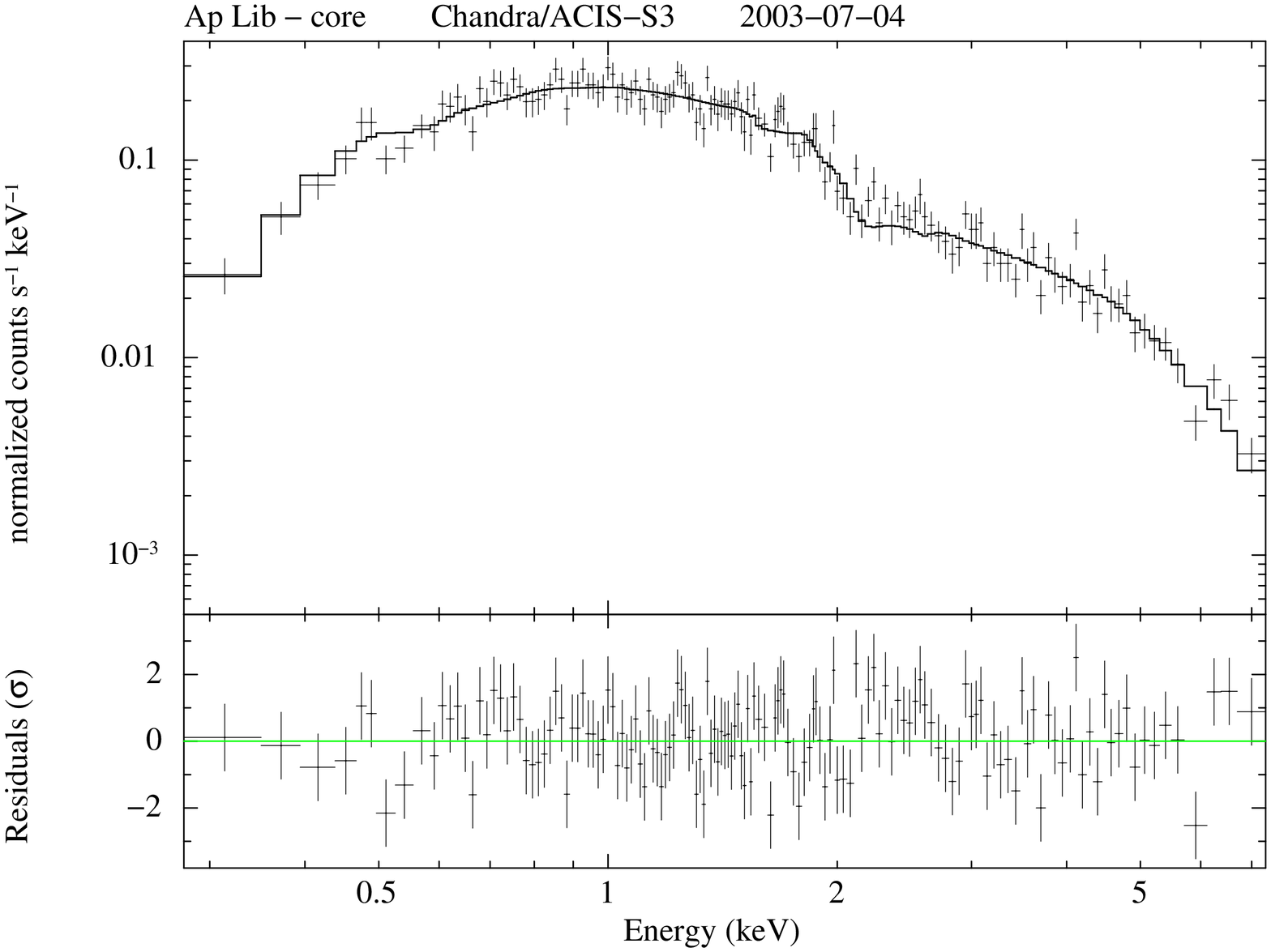}
              \vfil
\includegraphics[width=0.8\columnwidth]{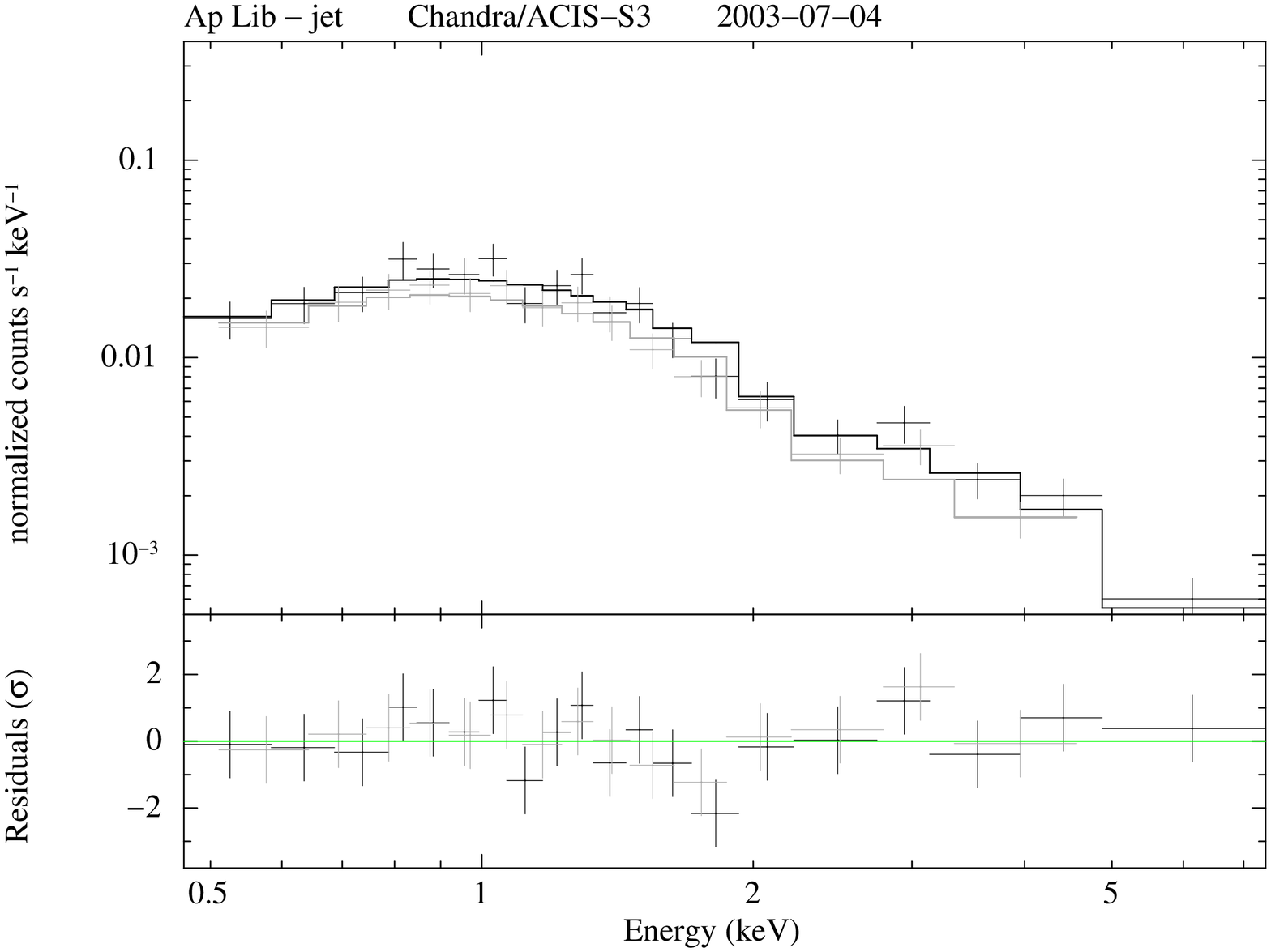}
   \caption{X-ray spectra extracted from the core and the jet.
The shown jet spectra are obtained from the {\it extended emission region} (black) and {\it wide jet region} (grey).
The line represent a power law fit taking into account the Galactic absorption as described in the text. The lower panels show the residuals of each fit. 
}
   \label{1514_chandra_spec}
 \end{figure}

The X-ray spectra for the core and the jet are shown in Fig. \ref{1514_chandra_spec}.
No significant difference of the slope between the jet and the core spectrum was determined.
The jet spectrum can be well described with a power law model and no emission or absorption features were detected.  
A fit with the thermal model {\tt apec} in {\tt Xspec}, considering the Galactic absorption, resulted in a worse fit and a high gas temperature of $kT = 6 \pm 3 \; \rm{keV}$. 
Any acceptable fit of a combined model using the combination of the thermal model {\tt apec} \citep{Smith2001} and a power law is dominated by the power law. 
Therefore the  power law model is the favoured description for the X-ray spectrum of the jet and hence the jet is considered to be dominated by non-thermal emission.

The core and the jet spectrum have both photon indices of $\Gamma < 2$ and are thus interpreted to be inverse Compton dominated. 
The core spectrum is comparable with the original definition of AP Lib being a low-energy peaked BL Lac object with IC dominance in the X-ray spectrum.

The luminosity of the X-ray core is $L_{\rm{core,2-10keV}} = (1.53 \pm 0.05)\times 10^{43} \rm{erg \; s^{-1}}$
and the jet is $L_{\rm{jet,2-10keV}} = (5.6 \pm 0.7)\times 10^{41} \rm{erg \; s^{-1}}$.


\subsection{Variability of the X-ray emission}

The {\it Chandra} light curve of the  core region of AP Lib has been created by extracting the counts from the source region and the background region used for the spectral analysis. The energy range of 0.2 to 8 keV was used.
The light curve has been studied in different binning, optimized for the exposure and frame time ($80.08\; \rm{s}$, $160.165\; \rm{s}$, $320.33\; \rm{s}$, $200\; \rm{s}$ , $1000\; \rm{s}$). 
A periodic signal was found in the extracted light curve, that can be explained by the dithering of {\it Chandra}. 
The dither period for the ACIS detector in the X direction is 1000s (Y direction 707s) .
As mentioned in \footnote{http://cxc.harvard.edu/ciao4.4/why/dither.html}, the dither period becomes visible when the source pass a node boundary, bad pixels or the chip edges. 
In the case of AP Lib, the node between the detector coordinate in X (CHIPX) of 511 to 512 is located in the dither direction. This cause lower count rates with a period of 1000 seconds. 
Except for this instrumental effect, the X-ray emission of the core is not variable.
The fit of a constant to the light curve with a binning of $\sim 200 \; \rm{s}$ 
results in an average count rate of $0.497 \pm 0.007 \; \rm{counts/sec}$ and a fit probability of $p_{\chi^2} \sim 30\%$ ($\chi^2 / dof = 61/65$) and therefore no significant variation could be detected. 


On longer time scales, variation of the X-ray emission was determined with {\it Swift} observations conducted between 2007 and 2011. 
10 {\it Swift} observations (obsID: 00036341001 to 00036341010) were conducted in 2007, 2008, 2010 and 2011 with a total exposure of 28.2 ks.
The observations on May 14, 2007 (00036341002) and Feb. 16, 2010 (00036341009) were not taken into account due to their low exposure. 
For the {\it Swift} analysis, XRT exposure maps were generated with the {\tt xrtpipeline} to account for some bad CCD columns that are masked out on-board. The masked hot columns appeared when the XRT CCD was hit by a micro meteoroid.
Spectra of the {\it Swift} data in PC-mode have been extracted with {\tt xselect} from a circular region with a radius of $20 \; \rm{pixel} \approx 0.8'$ at the position of AP Lib, which contains $90\%$ of the PSF at 1.5~keV. 
The background was extracted from a circular region  with radius of $80 \; \rm{pixel} \approx 3'$ near the source. 
The auxiliary response files were created with {\tt xrtmkarf} and the response matrices were taken from the {\it Swift} package of the calibration database {\tt caldb}.

The flux in the energy range 2 - 10 keV has two different level of $F_{\rm{2-10 keV}} = (3.2 \pm 0.4) \times 10^{-12} \;\rm{erg \; cm^{-2} \; s^{-1}}$
in 2007/2008 and $F_{\rm{2-10 keV}} = (4.9 \pm 0.5) \times 10^{-12} \;\rm{erg \; cm^{-2} \; s^{-1}}$ 
in 2010/2011. 
A fit of a constant results in a flux of $F_{\rm{2-10 keV}} = (3.9 \pm 0.2) \times 10^{-12} \;\rm{erg \; cm^{-2} \; s^{-1}}$ and a probability of $p_{\chi^2} = 6.6 \times 10^{-5}$ . 
During the full time period of the {\it Swift} observations no spectral change appeared (power law with average photon index of $\Gamma = 1.6 \pm 0.1$ taking into account the Galactic absorption).
To check for short term variability, light curves with a binning of 200 s from each single observation were created and no significant variation was detected on this short time scales. 

The variation is assumed to result from the core region; the jet cannot be resolved in the {\it Swift} XRT observations and the core dominates the measured flux.

In the sum of the available {\it Swift} observations in photon counting mode (total exposure of $\sim 28 \; \rm{ks}$), the X-ray jet is not visible. 
Since the angular resolution of {\it Swift} is $18''$ compared to $1''$ of {\it Chandra}, AP Lib appears point-like and the {\it Chandra} jet is fully contained in the XRT PSF. Only possible extension at larger scale of the X-ray jet, which could not be seen with {\it Chandra} due to the used subarray, could be determined with {\it Swift}. But the exposure of the available {\it Swift} observation is too low to determine any very faint extension of the jet beyond $18''$ distance to the core.

\section{Multi-wavelength observations}

\subsection{Jet synchrotron emission}

\subsubsection{Radio emission}
\label{section_radio}

Observations with the Very Large Array (VLA) on AP Lib \citep{Cassaro1999} show the clear detection of the radio jet (see Fig.~\ref{1514_Xjet_VLA}). In VLA observations at 1.36 GHz and 4.88 GHz, the radio jet emerges along the south-east direction and bends towards north-east after $\sim 12''$, for a total extent of $\sim 55''$ \citep{Cassaro1999}. 
The flux density of the jet at 1.36 GHz is given as $210 \; \rm{mJy}$ and the core has a flux density of $\sim 1.6 \; \rm{Jy}$ \citep{Cassaro1999}.
The observations with the D array at 1.4 GHz show a diffuse emission on arcmin scale on the same side of the jet (\cite{Condon1998, Cassaro1999}). 
Instead, the jet on milli-arcsec scale emerges to the south with a position angle of $\sim 180^\circ$ (the VLBA monitoring program MOJAVE\footnote{http://www.physics.purdue.edu/MOJAVE/}, \cite{Lister2009}). The jet on arcsec scales determined with VLA is pointing at PA $\sim 120^\circ$. 
The further bend with an angle of $\sim 50^\circ$ result in a total change of direction of $\sim 110^\circ$ compared to the milli-arcsec scale jet.

The comparison of the kpc jet in radio and X-rays reveals the same jet morphology of the emission along the SE direction (see Fig.~\ref{1514_Xjet_VLA}). 
An IC emission model for the X-ray jet assumes that radio synchrotron radiation is emitted by the same electrons that give rise to the X-ray radiation via Compton scattering. This is supported independently by the X-ray spectral data and the similarity of the jet morphology in the radio and X-ray bands.

\subsubsection{Optical emission}
\label{1514opticalemission}

In previous works, there have been no indication for a 
jet in the high energy synchrotron range. In HST observations, the elliptical host galaxy is seen, but no significant excess emission associated with the X-ray/radio jets is visible.

The core properties were studied by \cite{Westerlund1982} who used UBV measurements to separate a nucleus with a non-thermal continuum spectrum and an extended component identified as an elliptical galaxy.
The optical flux of the unresolved nucleus varied between 14.6 and 17 mag in the V-band during the observations mentioned in \cite{Westerlund1982} and the colours are given as $(B-V) = +0.55 \pm 0.01$ and $(U-B) = -0.57 \pm 0.01$. For the extended component they extracted an optical flux of $14.7 \pm 0.05$ mag in the V-band after correction for reddening and determined the color of the host galaxy as $(B-V)_{\rm galaxy} = 1.02$.

In 1993, \cite{Stickel1993} found that the host galaxy of AP Lib appears asymmetric and elongated towards a nearby galaxy ($\approx 65''$ to the north east). They suggested that AP Lib is an interacting system.
The obtained spectra by \cite{Pesce1994} show several absorption lines for the second galaxy resulting in a redshift of $z = 0.048$. This indicates that the host galaxy of the BL Lac object and the nearby galaxy are associated and the projected separation is 83 kpc \citep{Pesce1994}.


\subsection{Core and jet spectral distribution}

\begin{figure}[h]
\includegraphics[width=0.6\columnwidth]{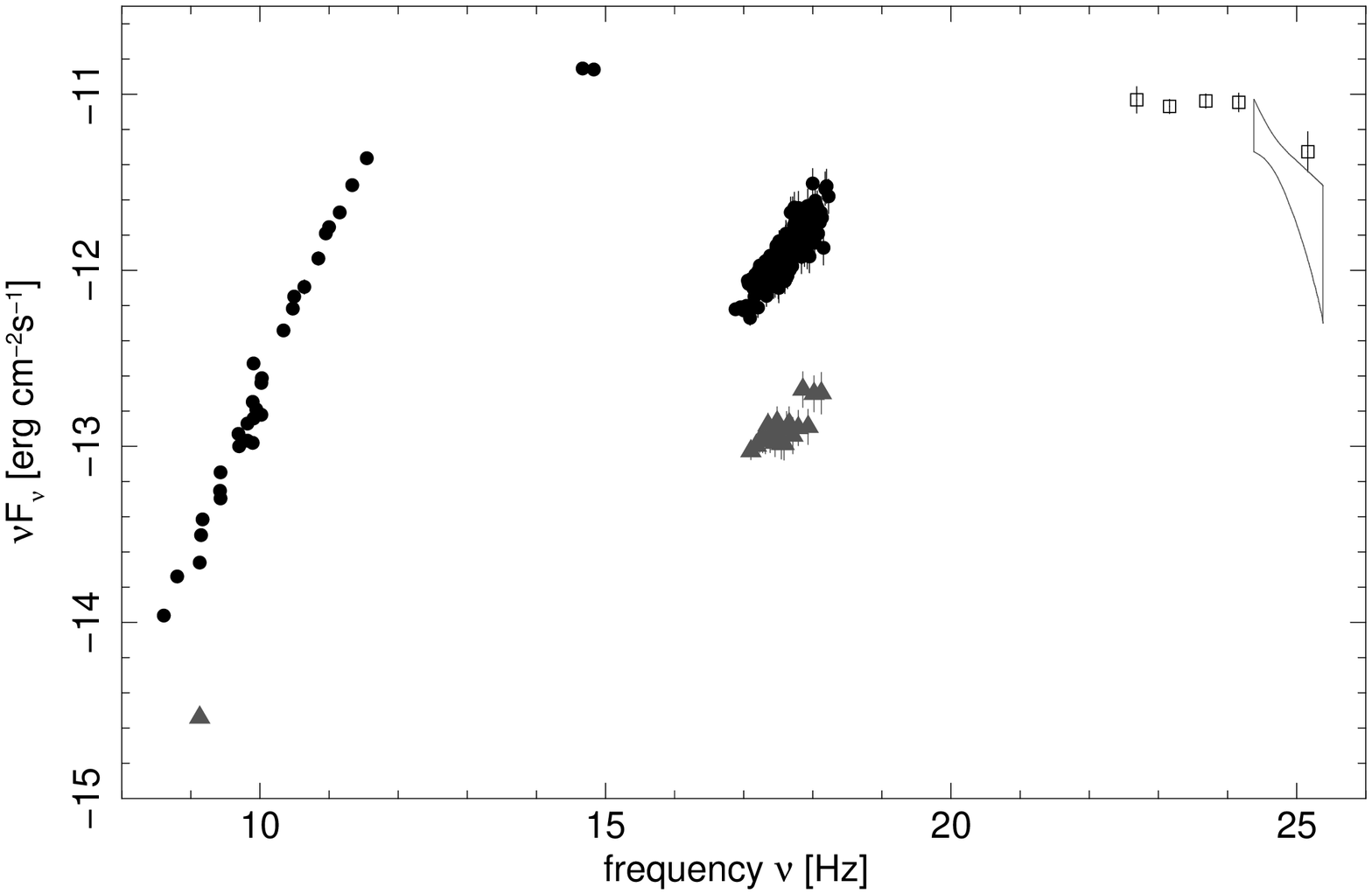}
   \caption{Historical radio (\cite{Kuehr1981,Cassaro1999} and PLANCK), optical (ATOM) 
and X-ray spectra ({\it Chandra}) of the core (black) of AP Librae. The optical fluxes are corrected for the influence of the host galaxy and the Galactic extinction and the X-ray spectra are corrected for Galactic absorption.
In grey triangles, the radio \citep{Cassaro1999} and X-ray spectra ({\it Chandra}) of the jet are shown.
 At $\gamma$-ray energies, the spectra from the 2FGL (open squares, \cite{Nolan2012}) and the 1FHL (butterfly, \cite{Ackermann2013}) catalogues are shown.
At these energies, no discrimination between jet and core emission is possible.
}
   \label{1514_corejetSED}
 \end{figure}

The spectra obtained from the {\it Chandra} observation for the core and jet region as well as radio and optical fluxes are summarized in Figure \ref{1514_corejetSED}. For the core distribution, the radio core flux from \cite{Cassaro1999} and the unresolved radio emission from PLANCK and \cite{Kuehr1981} dominated by the core emission, are shown.
The optical emission in the R and B band are values averaged over the time range of the 2FGL catalogue \citep{Nolan2012} obtained with the ATOM telescope. It has to be considered that AP Lib is very variable in the optical band. The optical emission has been corrected for the influence of the host galaxy \citep{Pursimo2002} and the Galactic extinction (using $E(B-V)=0.138$, \cite{Schlegel1998}). 
The radio emission from the jet is taken from \cite{Cassaro1999}. 
The spectral shape  of the core and jet emission is rather similar and the radio and optical emission can be described by synchrotron photons. Instead, the X-ray spectra are clearly dominated by inverse Compton emission due to their hard spectral index.

The information for the spectral energy distribution of the extended emission is still limited and is difficult to derive a detailed emission model. The hard spectral index of the X-ray jet suggests that the jet is dominated by inverse Compton emission in a similar way as the core.

 The high energy $\gamma$-ray spectrum from the 2FGL \citep{Nolan2012} and the 1FHL \citep{Ackermann2013} catalogues are also shown in Figure  \ref{1514_corejetSED}. It is not possible to discriminate between jet and core emission, due to the limited angular resolution of the {\it Fermi}-LAT instrument. 
In the 2FGL catalogue, the data are obtained over two years and no strong variation was found for AP Lib. The 1FHL catalogue considers one additional year of {\it Fermi}-LAT  observations, and is therefore not contemporaneous with the 2FGL.
The butterfly spectrum for the 1FHL data represents the $1\sigma$ uncertainty range of the flux density and photon index.


\subsection{High energy IC emission}

The instruments which can be used to detect the high energy IC emission above 100 MeV do not have the angular resolution to discriminate between jet and core emission. 

A GeV source, 2FGL J1517.7-2421 \citep{Nolan2012} was detected with the {\it Fermi} Gamma-ray space telescope and can be associated with AP Lib. 
The flat spectrum and its flux in the {\it Fermi} energy range indicates that the maximum of the high energy component of the emission is located at $> 100 \; \rm{MeV}$.


Very high energy (E$>$ 100 GeV) $\gamma$-ray emission was detected from AP Lib with the H.E.S.S. Cherenkov telescope array in June/July 2010 \citep{Hofmann2010}. 

Among all TeV BL Lac objects, AP Lib is the object with the lowest synchrotron peak frequency and the most extreme spectral indices for the radio-optical and optical-X-ray range \citep{Fortin2011}. 
Following the standard SSC model, the expected emission for AP Lib in the very high energy band is well below the sensitivity of current Cherenkov telescopes and it is remarkable to detect TeV emission from this object.

\section{Summary and conclusion}

AP Lib is classified as a low frequency BL Lac object. 
Unexpected for this subclass, the object has been detected in the TeV band \citep{Hofmann2010,Fortin2011}.
Generally, the emission model for explaining the TeV $\gamma$-ray emission at the base of a synchrotron self-Compton (SSC) model requires rather high Doppler factors.
The values found in spectral modelling significantly exceed the values determined from VLBI monitoring. 

AP Lib shows the lowest peak frequency of the synchrotron emission and the most extreme spectral indices of all TeV BL Lac objects.  This raises the question whether the jet properties of AP Lib are unusual as well.

A clearly visible extended, non-thermal X-ray jet was discovered in AP Lib in the analysis of the {\it Chandra}  observation of July 4, 2003. 
The X-ray jet is located in the south-east direction of the source and bends towards north-east comparable with the jet visible in the VLA radio observation. 
The radial profile of the X-ray jet reveals an extension up to $\sim 15''$, while no counter jet could be detected.
The detailed study of different profiles of the X-ray jet show that the X-ray emission morphology is comparable with the radio emission at 1.36 GHz. Therefore it is reasonable to assume, that the same electron population produces the radio and X-ray emission in the jet.
The X-ray and radio jet emerges in south-east direction and turns by an angle of $\sim 50^\circ$ 
between $2''$ and $20''$ and by $110^\circ$ between few milli-arcseconds and few arcseconds. 
This strong bending of the jet could be explained by a small angle of the jet axis to the line of sight of the observer. This would match the assumption of a high Doppler factor as expected from the detection of the TeV emission for which beaming is necessary. 
\cite{Rector2003} concluded that LBL are seen closer to the jet axis than HBLs since they have a wide distribution of parsec- and kpc-scale jet alignment angles. 

BL Lac objects represent FRI galaxies in which the jet points under a small angle of $\theta < 15^\circ$ \citep{Marscher2011} to the line of sight. Considering this maximum angle, the de-projected length ($d_{\rm{jet}}$) of the visible X-ray jet would be $d_{\rm{jet}} \approx 54 \; \rm{kpc}$. The possible range for the de-projected length would be up to $d_{\rm{jet}} \approx 802 \; \rm{kpc}$ assuming a much smaller angle of $\theta \approx 1^\circ$.
This maximum value for the de-projected length is comparable to the value determined by \cite{Marscher2011} for OJ287 of $d_{\rm{jet}} \approx 640 \; \rm{kpc}$ for $\theta = 3.2$.

The X-ray spectrum of the core region is described by a power law with $\Gamma = 1.58 \pm 0.04$ taking into account the Galactic absorption of $N_H = 8.36 \times 10^{20} \; \rm{cm^{-2}}$.
The jet spectrum 
is described by a power law with photon index of $\Gamma = 1.8 \pm 0.1$ and has a flux of $\sim 10\%$ of the core flux. 

Interestingly the core and jet spectra both have a hard spectral index indicating IC dominance. 
As described in e.g. \cite{Harris2006}, BL Lac objects are interpreted as beamed versions of the low-luminosity FRI radio galaxies. The IC dominance of the X-ray jet is unusual for low luminosity AGN, as described in \cite{Harris2006} and \cite{Worrall2009}. 
Generally, the X-ray jet spectra for FRI 
are synchrotron dominated and IC spectra only appear in X-ray jets of high luminosity sources. 


All six BL Lac objects with extended X-ray jets (AP Lib (this work), S5 2007+777 \citep{Sambruna2008}, OJ287 \citep{Marscher2011}, PKS 0521-365 \citep{Birkinshaw2002}, 3C371 and PKS 2201+044  \citep{Sambruna2007}) show jet luminosities intermediate between FRI and FRII. 

The spectral shape of the X-ray emission of their jets is comparable to be flat or IC dominated. AP Lib, OJ 287 and S5 2007+777 show stronger IC dominance. Therefore the BL Lac objects behave more similar to FRII which have mostly IC dominated X-ray jets, although BL Lac objects are generally classified as beamed versions of FRI galaxies. 
The X-ray jets of BL Lac objects are also more luminous than the X-ray jets of FRI galaxies. Compared to the sample of FRI X-ray jets of \cite{Harwood2012}, the X-ray luminosity of the jets of 3C371, PKS 2201+044 \citep{Harwood2012} and PKS 0521-365 \citep{Birkinshaw2002} are at the high end of the luminosity distribution of FRI with $ 2\times 10^{16} \; \rm{W/Hz} < L_\nu < 3\times 10^{17} \; \rm{W/Hz}$. Their X-ray spectra are rather flat and the photon indices are comparable to $2$. The jet of OJ 287, S5 2007+777 \citep{Harwood2012} and AP Lib (this work) have highest luminosities of $ 6\times 10^{17} \; \rm{W/Hz} < L_\nu < 8\times 10^{18} \; \rm{W/Hz}$ and IC dominated X-ray spectra. 
The X-ray jet spectra become IC dominated with increasing luminosity.

Within the group of BL Lac objects with X-ray jets, the spectral indices of the jets also correlate with the spectral indices of the cores. 
The LBL have hard core spectra and show also harder spectra in the jet. The LBLs S5 2007+777 \citep{Sambruna2008} and OJ287 \citep{Marscher2011} have, like AP Lib, X-ray jets with IC dominated spectra. 
The IBL have softer spectra in the core 
and their jet spectra also appear softer than the ones for LBLs. 
Hence, BL Lac objects with low synchrotron peak energies have higher probability to have IC dominance in the X-ray jet.
AP Lib has the lowest luminosity in the X-ray jet among the jets of BL Lac objects with a clear IC dominated X-ray spectrum.
This raises the question, if the explanation for IC jets as IC scattering of CMB photons is still working or whether other external photon fields become important. 

Since also the core components of the X-ray emission of the BL Lac objects are IC dominated (only the core emission of S5 2007+777 has $\Gamma \sim 2$), this component is expected to reach high energies. 
Furthermore, the IC spectrum emerges in the X-ray regime of the core emission and the fact that the peak of the IC emission is in the GeV range (all six BL Lac objects are detected by {\it Fermi} have $2.0 < \Gamma < 2.4$ in the 2nd {\it Fermi} catalog \citep{Nolan2012}) reveal a broad IC component.
This behaviour of the IC scattering could also exist in the jet component. For AP Lib it is already known that the IC component reaches TeV energies \citep{Hofmann2010}.

The discovery of an extended X-ray jet in the TeV BL Lac object AP Lib, which is IC dominated, provides therefore the opportunity to test the interpretation of the most common emission models as summarized in the review of X-ray jets, e.g. \cite{Harris2006}.



\acknowledgments
The scientific results reported in this article are based on data obtained from the {\it Chandra} Data Archive. This research has made use of software provided by the {\it Chandra} X-ray Center (CXC) in the application package CIAO.
The use of the public HEASARC software packages is also acknowledged. 
S.K. and S.W. acknowledge support from the BMBF through grant DLR 50OR0906.


\end{document}